\documentclass[preprint,amsmath,amssymb,showpacs,english,aps]{revtex4}
\usepackage{graphicx}
\makeatletter
\usepackage{times}
\usepackage{subfigure}
\usepackage{amssymb}
\usepackage{amsmath}

\makeatletter
\numberwithin{equation}{section} 
\newtheorem{theorem}{Theorem}
\newtheorem{proposition}[theorem]{Proposition}
\newcommand{\be}{\begin{equation}}
\newcommand{\ee}{\end{equation}}
\newcommand{\ben}{\begin{eqnarray}}
\newcommand{\een}{\end{eqnarray}}

\newcommand{\nd}{\noindent}

\usepackage{babel}
\makeatother
\begin{document}

\title{Scale invariance and related properties of  q-Gaussian systems}

\author{$^1$C. Vignat and A. Plastino$^2$}
\address{$^1$ L.P.M., E.P.F.L, Lausanne, Switzerland \\
$^2$ La Plata National University, Exact Sciences Faculty\\ $\&$
National Research Council (CONICET) \\ C. C. 727 - 1900 La Plata -
Argentina  }

\begin{abstract}
We advance scale-invariance arguments for  systems that are
governed (or approximated) by a $q-$Gaussian distribution, i.e., a
power law distribution with exponent $Q=1/(1-q);\,\,\,q \in
\mathbb{R}$. The ensuing line of reasoning is then compared with
that applying for Gaussian distributions, with emphasis on
dimensional considerations. In particular, a Gaussian system may
be part of a larger system that is not Gaussian, but, if the
larger system is spherically invariant, then it is necessarily
Gaussian again.
  We show that this result extends to q-Gaussian systems via elliptic
  invariance.
The  problem of estimating the appropriate value for $q$ is
revisited. A kinetic application is also provided.

 \vskip 0.5 mm \nd 

 \vskip 0.5 mm \nd {\bf
Keywords}: Scale invariance, q-Gaussian
 distributions.

\end{abstract}

\maketitle

\section{Introduction and background material}

\nd Homogeneous power-laws, such as Newton's universal law of
gravitational attraction, for instance, abound in Nature. They
are, by definition, self-similar and thus true in all scales.
Systems statistically described by {\it power-law probability
distributions}
 are rather ubiquitous \cite{cero} and thus
of perennial interest \cite{uno}. In this report we wish to give
careful scale-invariance consideration to systems that are
governed (or described) by a special kind of power-law probability
distribution functions (PDF), namely, the $q-$Gaussian function.
Consider a system $\mathcal{S}$ described  by a vector $X$ with
$n$ components. We say that $X$ is $q-$Gaussian distributed if its
probability distribution function writes as described by
(\ref{eq:q>1gaussian})-(\ref{eq:Kq>1}) below.

 It is well-known that for such $\mathcal{S}-$systems one can appeal to
 Jaynes' maximum entropy principle (MaxEnt) \cite{dos} under $\,$
 a covariance constraint with
 a generalized (or $q-$) information measure
\be \label{tsallis}
H_{q}\left(x\right)=(1-q)^{-1}\,\int\,dx\,[f(x)-f^{q}(x)];\,\,\,\,\,\,\,\,\,q
\in \mathcal{R}, \ee
 as the protagonist \cite{alfo}. This  measure has been
 found to be useful in extracting information pertaining to systems that are
 characterized by  either (1) fractal nature, (2) long-memory,
 or
 (3)  long-range interactions \cite{dino}. Employing $H_q$ for {\it
 other} types of system has generated controversy \cite{euro} which is of no
 relevance to our present purposes.   We will focus attention upon
   properties of Gaussian systems that remain valid for their
q-Gaussian counterparts as well (as  $q$ becomes different from
unity), with emphasis on the dimensional properties of both kinds
of  systems. It is well-known that if a system ``is Gaussian", any
part (sub-system) of it is still Gaussian. This property holds for
$q-$Gaussian systems as well, as proved in \cite{hero}. {\it A
more interesting result is the inverse phenomenon}: a Gaussian
system may be part of a larger system that is not Gaussian.
However, if the larger system is spherically invariant, then it is
necessarily Gaussian again. Surprisingly enough, this ``inverse
property" has gone largely ignored  in the statistical literature.
In this work we will not only provide a simple proof for it but we
will show that it can be  extended to q-Gaussian
 distributions as well.
These results can be given a physical interpretation within the
framework of the estimation of the  parameter $q$ of a given system
\cite{dino}. We will prove that, if spherical symmetry prevails,
such estimation can be performed using only a {\it restricted},
observable part of the system and that the  overall parameter $q$
for the entire system  can be retrieved provided the dimension of
the system is known. We begin our considerations by introducing
the two basic notions, namely spherical symmetry and $q-$Gaussian
systems. We will also apply our results to a simple case of
kinetic theory via the Beck-Cohen superstatistics theory
\cite{B1,B2,B3,B4,B5,B6,B7}.

\subsection{Spherical symmetry}
A really momentous symmetry is that of  invariance against
rotations. It is found in the fundamental laws of nature and
constitutes one of the most powerful principles in elucidating the
structure of individual atoms, complicated molecules, and entire
crystals.   Also, it characterizes the shape of many systems. We
can cite  self-gravitating systems like stars and planets, that
have quasi-spherical shape if their mass is large enough. Also,
many atomic nuclei are spherical, and many molecules as well, etc.
Conservation of angular momentum, a very frequent occurrence, is a
result of the isotropy of space itself \cite{margenau}.  We
discuss now some properties of spherical probability
distributions.

The characteristic function associated with a random vector $X \in
\mathbb{R}^n$ is \be \label{uno} \varphi_X(U)=  E e^{iU^tX};\,\,\,
U \in \mathbb{R}^{n}. \ee Under the hypothesis, discussed for
instance in the textbook  \cite[XV.3]{feller}, that $ \varphi_{X}
\in \mathcal{L}_{1}(\mathbb{R}^n)$, there is a one-to-one relation
between $\varphi_{X}$ and the probability density function
$f_{X}$ of $X$. The random vector $X$ is said to have a spherical
distribution if its characteristic function $\varphi_X$ satisfies
\begin{equation}
\varphi_X(U)=\phi(\Vert U \Vert)
\end{equation}
for some scalar function $\phi : \mathbb{R}^+ \rightarrow \mathbb{R}$ which is
then called the characteristic generator of the spherical
distribution.
We will write $X \sim \mathcal{S}_n(\phi)$ in this case.
It is well-known that an equivalent definition for a spherical random vector $X$ is
\[
X \sim AX; \,\, \forall \, A \,\, \text{orthogonal}
\] where $\sim$ denotes equality in distribution.

Spherical random vectors have, as it is well-known,  the following
properties:
\begin{enumerate}
\item All marginal distributions of a spherical distributed random vector
are spherical.
\item All marginal characteristic functions have
the same 
characteristic generator.
\item If  $X \sim \mathcal{S}_n(\phi)$ then
\begin{equation}
X \sim rT_{n}
\end{equation}
where $T_{n}$ is a random
vector distributed uniformly on the unit sphere surface in
$\mathbb{R}^n$  and $r$
is a positive random variable independent of
$T_{n}$.
\end{enumerate}
Let us remark that a spherically distributed random vector does
not necessarily possess a density.

A generalization of the concept of spherical distribution is given
by elliptical distributions, to which the multi-normal
distribution belongs. Elliptical distributions have recently
gained a lot of attention in financial mathematics, being of use
particularly in risk management. A  $n-$random vector $Y$ is said
to have an elliptical distribution with so-called characteristic
matrix $C_Y (n \times n)$ if $Y \sim AX$, where $X \sim
\mathcal{S}_n(\phi)$ and $A$ is a $n \times n$ deterministic
matrix such that  $A^tA=C_Y$ and $rank(C_{Y})=n$. We shall write
$Y \sim \mathcal{E}_{n}(C_Y,\phi)$.

\subsection{$q-$Gaussian systems}
An $n-$components vector $X$ is $q-$Gaussian distributed if its
PDF writes as follows \cite{uno}:
\begin{itemize}
\item
in the case $1<q<\frac{n+4}{n+2}$
\begin{equation}
f_{X}\left(X\right)=A_{q}\left(1+X^{t}
\Lambda^{-1}X\right)^{\frac{1}{1-q}},\label{eq:q>1gaussian}
\end{equation}
matrix $\Lambda$ being related to the covariance matrix
$K=EXX^{t}$ in the fashion  \cite{vignat1}
\begin{equation}
\Lambda=\left(m-2\right)K.\label{eq:Kq>1}
\end{equation}
 where the 
number of degrees of freedom is defined as  \cite{vignat1}
\begin{equation}
m=\frac{2}{q-1}-n.
\label{eq:mq>1}
\end{equation}
Moreover, the partition function $Z_q= 1/A_q$ reads
\cite{vignat1}
\[
Z_{q}=\frac{\Gamma\left(\frac{1}{q-1}-
\frac{n}{2}\right)\vert\pi\Lambda\vert^{1/2}}{\Gamma\left(\frac{1}{q-1}\right)}.
\]
and the characteristic function is
\begin{equation}
\varphi_{X}(U)
= \frac{2^{1-\frac{m}{2}}}{\Gamma(\frac{m}{2})}z^{\frac{m}{2}}K_{\frac{m}{2}}(z)
\end{equation}
with $z=\sqrt{U^t\Lambda U}$ and $K$ is the modified Bessel function of the second kind.

\item
in the case $q<1$
\begin{equation}
f_{X}\left(X\right)=A_{q}\left(1-X^{t}
\Sigma^{-1}X\right)_{+}^{\frac{1}{1-q}}\label{eq:q<1gaussian}
\end{equation}
with matrix $\Sigma=dK$ and parameter $d$ defined as
$d=2\frac{2-q}{1-q}+n.$ In this case, the partition function is
\[
Z_{q}=\frac{\Gamma\left(\frac{2-q}{1-q}\right)
\vert\pi\Sigma\vert^{1/2}}{\Gamma\left(\frac{2-q}{q-1}+\frac{n}{2}\right)}.
\]
and the characteristic function is
\begin{equation}
\label{chfq>1} \varphi_{X}(U) = 2^{\frac{d}{2}-1}
\Gamma(\frac{d}{2})\frac{J_{\frac{d}{2}-1}(z)}{z^{{\frac{d}{2}-1}}}
\end{equation}
where $z=\sqrt{U^t\Sigma U}$ and $J$ is the Bessel
function of the first kind.
\end{itemize}
We begin in the next Section to advance our present results.

\section{Size behavior of the $\,$ $q\,-$parameter }

Our  first result revolves around   the behavior of the
non-extensivity parameter $q$ as a function of the dimension of
the system and is embodied in the following theorem, the proof
of which is given in the Appendix.

\begin{theorem}
\label{thm1} Assume that a system $X_n\in\mathbb{R}^{n}$ follows a $q-$Gaussian
distribution with parameter
$q_{n}$; then with $1\le k\le n-1,$ any $k-$dimensional subsystem
$X_{k}=\left[x_{1},\dots,x_{k}\right]^{t}$  of $X_n$ is $q-$Gaussian distributed with parameter

\begin{equation}
\label{qk}
q_{k}=1-\frac{2\left(1-q_{n}\right)}{2+\left(n-k\right)\left(1-q_{n}\right)}
\end{equation}

Reciprocally, assume that a $n-$dimensional  and spherical system
$X_n$ contains a $k-$dimensional subsystem $X_{k}$ that follows a
$q-$Gaussian distribution with  parameter $q_{k}$; then system $X_n$
is itself $q-$Gaussian distributed with parameter $q_{n}$ defined
as in (\ref{qk})
\end{theorem}

Recall that 
for any
$n-$dimensional orthogonal transformation $A$, there exists an
orthogonal decomposition of $\mathbb{R}^n$ as
\[
\mathbb{R}^{n}=E \oplus F \oplus G_1 \oplus \dots \oplus G_k
\]
into stable subspaces, the restriction of $A$ to each subspace being
\begin{itemize}
\item the identity transformation for subspace $E$
\item minus the identity transformation for subspace $F$
\item a two-dimensional 
planar rotation of angle $\theta_k$ for subspace $G_k$
\end{itemize}
Moreover, $n-$dimensional
$q-$Gaussian distributions with parameter $q_{n}$ arise in
statistical physics as the canonical distributions of systems with
maximal $q-$entropy $H_q$ of order $q_{n}$
and fixed covariance matrix.  The result of theorem
(\ref{thm1}) can be paraphrased in this way: if a system of
dimension $k$ has maximum $q-$entropy of order $q_{k}$, and is
also  part of a spherical system of dimension $n>k$, then the
whole system maximizes the $q-$entropy with parameter $q_{n}$
related to $q_{k}$ as in (\ref{qk}). Notice that in (\ref{qk}), as
$q_{k}\rightarrow1$ then $q_{n}\rightarrow1$ and we deduce that if
a spherical system has a Gaussian part, then it is
necessarily Gaussian as well.

We note  also that $q_{n}<1$ implies $q_{k}<1$ while $q_{n}>1$
implies $q_{k}>1.$ This result is natural since cases $q>1$ and
$q<1$ correspond to two different types of distributions:
according to Beck and Cohen's superstatistics principle
\cite{B1,B2,B3,B4,B5,B6,B7}:
\begin{itemize}
\item  $q>1$ that
corresponds to a Gaussian system subjected to fluctuations that
are independent of the state of the system.
\item  $q<1$ that
corresponds again to a fluctuating Gaussian system for which the
amplitude of the fluctuations depends on the  system's state.
\end{itemize}
Alternatively, these two cases can be characterized as follows.
Our system is here described by the random vector $X$: if $q>1$,
then $X$ has unbounded support, contrarily to the bounded support
associated to the case $q<1$ \cite{vignat1}.

Thus, we obtain the following rather  natural result: {\it
reduction or enlargement of a $q-$Gaussian system does not change
its superstatistical nature}.

\section{A second result: multi-component systems}

\subsection{Average behaviour}

It may happen that  measuring the behavior of only one component
$x_{1}$ of a large system $X \in \mathbb{R}^n$ is not physically
feasible, and one has to content oneself with measuring instead
the behavior of a superposition of contributions from  (or average
of)  several components \cite{roybook}, in the form

\begin{equation}
\label{linearcombination}
<X> = \sum_{i=1}^{n}a_{i}x_{i}
\end{equation}
where the deterministic coefficients  $a_{i} \in \mathbb{R}$
characterize the measurement device. The following theorem (see
Ref. \cite{fang}) allows to give a special characterization of
this average value in the case of spherical systems.

\begin{theorem}
\label{thm2} \cite[Th. 2.4]{fang} If $X\in\mathbb{R}^n$  is
spherically distributed and $A=(a_1,\dots,a_n)^t$  is a deterministic
vector then
\[
<X> = \sum_{i=1}^{n}a_{i}x_{i}
\]
is distributed as $\Vert A \Vert x_{1},$ where $\Vert A \Vert$ is the Euclidean norm of $A.$
\end{theorem}
In the next subsection we extract rather interesting physical
conclusions from this theorem.

\subsection{Application to the estimation of $q$}

An important problem in non-extensive statistics is the estimation
of the non-extensivity parameter $q_{n}$ associated to an
$n-$dimensional system that follows a $q-$Gaussian distribution
\cite{dino}. We provide here some hints  about a possible
estimation strategy in the case $q>1$, assuming that we have
access to averaged measures of the system of the type
(\ref{linearcombination}).

Assuming  $q_n>1,$ then if $X_n$ follows  distribution
(\ref{eq:q>1gaussian}),  the averaged measure $<X>$ is distributed
as
\[
f_{<X>}(x)=\frac{A_{q_{1}}}{\Vert A \Vert}(1+\frac{x^2}{\lambda
\Vert A \Vert^2})^{-\frac{m+1}{2}}
\]
with $m=2/(q_{1}-1)-1$ and $\lambda=\Lambda_{1,1}.$ As a consequence, a possible  estimation
strategy of parameter $q_{n}$ follows the three following steps:
\begin{enumerate}
\item since the ``tail-behavior" of the distribution of $f_{<X>}$ is
\[
f_{<X>}(x) \sim x^{-(m+1)},
\]
(where $\sim$ means here asymptotic equivalence), parameter $m$ can be estimated as the   L\'{e}vy exponent  of the
distribution of the average measure of the system \cite{note}
\item   the non-extensivity parameter $q_{1}$
 of $<X>$ can be computed using (\ref{eq:mq>1}) as
\[
q_{1}=\frac{m+3}{m+1}
\]
\item the non-extensivity
parameter  $q_{n}$ of the $n-$dimensional  system $X_n$ can in
turn be deduced using (\ref{qk}) as
\[
q_{n} =\frac{2-(n+1)(1-q_{1})}{2-(n-1)(1-q_{1})}
\]
\end{enumerate}

As a new result we find that  if the dimension $n$ of the system
is known, its  non-extensivity parameter $q_{n}>1$ can be
evaluated from any measurement of the type
(\ref{linearcombination}).

\subsection{A kinetic application}
\subsubsection{Theoretical framework}
Another application of the latter result  can be provided in the
context of the kinematics of collision events. We envision a
scenario in which attention is focused 
on the particles of a
system interacting with a heat bath (a fundamental problem in
thermodynamics). An elastic collision between (i) a system's
particle with momentum $P$, mass $M$, velocity $V$, and energy $E$
and (ii) a particle from the heat bath with momentum $p$, mass
$m$, velocity $v$, and energy $\epsilon$, verifies \cite{dunkel}

\begin{eqnarray*}
E + \epsilon & = \hat{E} + \hat{\epsilon} \\
P + p & = \hat{P} + \hat{p}
\end{eqnarray*}
where ``hats" refer  to quantities after the collision.
In the non-relativistic case, these quantities write
\begin{eqnarray*}
P &= MV, \hspace{1cm} p=mv \\
E & = \frac{\Vert P \Vert ^{2}}{2M}, \hspace{1cm} \epsilon = \frac{\Vert p \Vert ^{2}}{2m},
\end{eqnarray*}
where momenta are $3-$dimensional  quantities. These equations can
be solved as
\begin{eqnarray*}
\hat{P}(p,P) &= \left(\frac{2M}{M+m}\right)p + \left(\frac{M-m}{M+m}\right)P   \\
\hat{p}(p,P) &=  \left(\frac{m-M}{M+m}\right)p + \left(\frac{2m}{M+m}\right)P
\end{eqnarray*}

Assuming that $P$ and $p$ are  independent random
variables, we look for stationary distributions for $p$ and $P$,
that is, for probability density functions $f_{p}$  and $f_{P}$
such that if $p\sim f_{p}$ and $P\sim f_{P}$ then after the
collision, $\hat{p}\sim f_{p}$ and $\hat{P} \sim f_{P}$. An
obvious pair of stationary solutions is given \cite{dunkel} by the
independent Maxwell solutions
\begin{equation}
f_{P}(P) = \frac{1}{\left( 2\pi M k_{B}T \right)^{3/2}}
\exp\left({-\frac{\Vert P \Vert ^{2}}{2Mk_{B}T}}\right), \,\, f_{p}(p) =
\frac{1}{\left( 2\pi m k_{B}T \right)^{3/2}}
\exp\left({-\frac{\Vert p \Vert ^{2}}{2mk_{B}T}}\right).
\end{equation}

\subsubsection{The correlated scenario}

Suppose however  that the assumption of independence between
momenta $p$ and $P$ does not hold. Such  is the case, for example,
when the corresponding particles are subject to the same
fluctuations (an interpretation for this scenario is provided in
the following Subsection). In this instance we look for a
stationary joint distribution for $p$ and $P$, i.e., for  a
probability density function $f_{p,P}$ such that if $(p,P)\sim
f_{p,P}$ then, after the collision, $(\hat{p},\hat{P}) \sim
f_{p,P}$ again. 
We note that this in turn implies $\hat{p} \sim
f_{p}$ and $\hat{P} \sim f_{P}.$ 

We need here an  extension of Theorem \ref{thm2} as given below,
the proof of which can be found, for example, in \cite{chu}.

\begin{proposition}
If $X\sim \mathcal{E}_n(C_X,\phi)$  and $A$ is a full-rank 
$(n \times n)$ matrix
then $Y=AX \sim \mathcal{E}_n(C_Y,\phi)$ with
\[
C_Y=AC_XA^t.
\]
\end{proposition}
Now assume that 
\[
C_X = \left[\begin{array}{cc} m I_3 & 0_3\\ 0_3 &
M I_3\end{array}\right], \,\, A =\frac{1}{m+M} \left[\begin{array}{cc}
(m-M)I_3 & (2m)I_3\\ (2M)I_3 & (M-m)I_3\end{array}\right]
\]
where  $0_3$ denotes the $(3\times3)$ null matrix and $I_3$ the $(3\times3)$ identity matrix. Then 
\[
C_Y=AC_{X}A^t=C_X.
\]
We are now in a position to deduce the following result:

\begin{theorem}
If $\left(\begin{array}{c}p \\P\end{array}\right)  \sim
\mathcal{E}_6(C_X,\phi)$ with characteristic matrix $C_X$ as
above, then the momenta vector after the collision
$\left(\begin{array}{c}\hat{p} \\ \hat{P}\end{array}\right) \sim
\mathcal{E}_6(C_X,\phi).$ As a consequence, any elliptical joint
distribution with characteristic matrix $C_X$ is stationary.  In
particular, $p$ and $\hat{p}$ have the same distribution, as well
as $P$ and $\hat{P}.$
\end{theorem}

\subsubsection{Superstatistics at work}

A more physical interpretation can  be given to the preceding
result, using the notion of superstatistics \cite{B1}. 
We know from \cite{dunkel}
that a pair of independent Gaussian momenta are stationary for the
collision process. Now,
\begin{itemize}
\item if $p\sim \mathcal{N}_{3}(mk_{B}T)$ (the $3-$dimensional
Gaussian distribution with covariance matrix $mk_{B}T I_3$) - and
\item $P\sim \mathcal{N}_{3}(Mk_{B}T)$, then \item  $\hat{p}\sim
\mathcal{N}_{3}(mk_{B}T)$ and $\hat{P}\sim
\mathcal{N}_{3}(Mk_{B}T)$. \end{itemize} Since
\[
\hat{P}(p,P) = \left(\frac{2M}{M+m}\right)p +
\left(\frac{M-m}{M+m}\right)P,
\]
choosing any (dimensionless) random variable  $a$ independent of
both $p$ and $P$, and defining the new quantities $q=ap$, $Q=aP$,
and $\hat{Q}=a\hat{P}$, we deduce that

\[
\hat{Q}(q,Q) = \left(\frac{2M}{M+m}\right)q +
\left(\frac{M-m}{M+m}\right)Q,
\]
so that, obviously, the pair  $(q,Q)$ is another couple of momenta
whose distribution is stationary. Obviously, variables $q=ap$ and $Q=aP$
are not independent ones, since they share the same random factor
$a$ (unless $a$  is almost surely a constant, which reduces to the
Gaussian case).

As a special case,
 if $a$ follows an inverse chi-distribution  with  $m$ degrees of freedom,
 one immediately finds ~\cite{B1} that the random vector
   $X=\left(\begin{array}{c}p \\P\end{array}\right)$ follows a Tsallis-distribution
\begin{equation}
f_{X}\left(X\right)=A_{q}\left(1+X^{t}
\Lambda^{-1}X\right)^{\frac{1}{1-q}},
\end{equation}
with non-extensivity  index $q>1$ related to parameter $m$ as in
formula (\ref{eq:mq>1}) with $n=6$. We remark that the  random
variable $a$ can be interpreted, in such a  context, as
representing temperature's fluctuations, as shown by Beck and
Cohen~\cite{B1}. Thus, the
presence of temperature fluctuations
indicates that the momenta of the incoming colliding particles are
correlated. Conversely, if they are correlated, then temperature
fluctuations ensue. This scenario is a feasible one if the heat
bath is a finite one, which, in turn, establishes a natural
connection with an old result of Plastino and Plastino
\cite{gibbscano}.

\section{Conclusions}

In this work we have considered physical applications of  a
largely ignored  result of the statistical literature: {\it a
Gaussian system may be part of a larger system that is not
Gaussian. However, if this larger system is spherically invariant,
then it is necessarily Gaussian again}.

We have provided a simple proof for it and we have shown  that it
can be extended to $q-$Gaussian distributions as
well. Our results have been  given a physical interpretation
within the framework of the problem of estimation of the q-Gaussian
 parameter
$q$. Also, we applied them to a simple instance of kinetic theory
involving Beck and Cohen superstatistics \cite{B1}.

\section{Appendix: Proofs}

\subsection{Proof of theorem \ref{thm1}}
We give here a  simple proof of theorem (\ref{thm1}), the
principle of which has been kindly suggested to us by Pr. Wlodek Bryc
and Pr. Jacek Wesolowski. Assuming first that
$X_{k}=[x_1,\dots,x_k]^{t}$ is $q-$Gaussian with parameter
$q_{k}>1$, we deduce that, with $U_{k}=[u_{1},\dots,u_{k}]^t,$
\[
\varphi_{X_k}(U_k) =\phi(\Vert U_{k} \Vert) =
\frac{2^{1-\frac{m}{2}}}{\Gamma(\frac{m}{2})}\Vert U_{k}
\Vert^{\frac{m}{2}}K_{\frac{m}{2}}(\Vert U_{k} \Vert)
\]
so that
\[
\phi(u)= \frac{2^{1-\frac{m}{2}}}{\Gamma(\frac{m}{2})}\Vert u
\Vert ^{\frac{m}{2}}K_{\frac{m}{2}}(\Vert u \Vert)
\]
and
\[
\varphi_{X_n}(U_{n}) =\phi(\Vert U_{n}
\Vert)=\frac{2^{1-\frac{m}{2}}}{\Gamma(\frac{m}{2})}\Vert U_{n}
\Vert ^{\frac{m}{2}}K_{\frac{m}{2}}(\Vert U_{n} \Vert)
\]
and $X_{n}$ is $q-$Gaussian  with dimension $n$ and $m$ degrees of
freedom, and thus has non-extensivity parameter $q_{n}$ such that

\[
m=\frac{2}{q_{k}-1}-k=\frac{2}{q_{n}-1}-n .\] The same result
applies  with $q_{k}<1$ by considering characteristic function
defined as in (\ref{chfq>1}).

\subsection{An alternate proof of theorem \ref{thm1}}
We provide here  an alternate proof based on stochastic
representations, as first used in \cite{liang}, extending their
result to the case $q>1$. Assume that $X_{k} \in \mathbb{R}^{k}$
is $q-$Gaussian distributed with parameter $q_{k}$: a stochastic
representation of $X_{k}$ is  (see \cite{fang})
\[
X_{k}=\frac{\chi_{k}}{\chi_{d}}Z_{k}
\]
where $\chi_{k}$ and $\chi_{d}$ are independent and chi
distributed random variables,  where $d=\frac{2}{q_{k}-1}-k$ and
$Z_k$ is uniform on the sphere. We know moreover that if
$X_{n}=rZ_{n}$ then
\[ X_{k}=rd_{1}Z_{k}\] where
$d_{1}^{2}\sim\beta_{\frac{k}{2},\frac{n-k}{2}}.$ We deduce that\[
rd_{1}Z_{k}=\frac{\chi_{k}}{\chi_{d}}\] or\[
r^{2}d_{1}^{2}=\frac{\chi_{k}^{2}}{\chi_{d}^{2}}.\] But by Luckacs
theorem \cite{lucas}:

\[
\frac{\hat{\chi}_{n}^{2}}{\hat{\chi}_{d}^{2}}
\frac{\tilde{\chi}_{k}^{2}}{\tilde{\chi}_{k}^{2}
+\tilde{\chi}_{n-k}^{2}}=\frac{\chi_{k}^{2}}{\chi_{d}^{2}}\] we
deduce that \[ r=\frac{\chi_{n}}{\chi_{d}}\] and that
$X_{n}=rZ_{n}$ is $q-$Gaussian distributed with parameter $q_{n}=
1+\frac{2}{d+n}=$$1+\frac{2\left(q_{k}-1\right)}{2+\left(n-k\right)\left(q_{k}-1\right)}.$


\begin{thebibliography}{1}
\bibitem{note} Note
that there exist other statistical approaches to the estimation of
parameter $q_{1}$: see \cite{korolev}.
\bibitem{lucas} If $X$ and $Y$ are independent then $\frac{X}{X+Y}$ and $X+Y$  are
independent if and only if  they have standard Gamma
distributions. See \cite{Yor}.
\end{thebibliography}
\end{document}